# Light-Activated Motion, Geometry- and Confinement-Induced Optical Effects of 2D Platelets in a Nematic Liquid Crystal

Antonio Tavera-Vázquez[1], Danai Montalvan-Sorrosa[2,†], Gustavo Perez-Lemus[1],

Otilio E. Rodriguez-Lopez[3,4], Jose A. Martinez-Gonzalez[3],

Vinothan N. Manoharan[2,5], Juan J. de Pablo[1,6]*

[1]Pritzker School of Molecular Engineering, The University of Chicago, Chicago, IL 60637, USA.

[2]Harvard John A. Paulson School of Engineering and Applied Sciences,
 Harvard University, Cambridge, MA 02138, USA.

[3]Facultad de Ciencias, Universidad Autónoma de San Luis Potosí,
 Av. Parque Chapultepec 1570, San Luis Potosí, 78210 SLP, México.

[4]Instituto de Física, Universidad Autónoma de San Luis Potosí,
 Av. Parque Chapultepec 1570, San Luis Potosí, 78295 SLP, México.

[5]Department of Physics, Harvard University, Cambridge, MA 02138, USA.

[6]Materials Science Division, Argonne National Laboratory, Lemont, IL 60439, USA.

[†]Present Address: Instituto de Química, Universidad Nacional Autónoma de México,
 Circuito Exterior, Ciudad Universitaria, Alcadía Coyoacán, 04510, Ciudad de México, México.

*Corresponding author:
 Juan J. de Pablo.  E-mail: depablo@uchicago.edu

## Abstract

Motile liquid crystal (LC) colloids show peculiar behavior due to the high sensitivity to external stimuli driven by the LC elastic and surface effects. However, few studies focus on harnessing the LC phase transitions to propel colloidal inclusions by the nematic-isotropic (NI) interface. We engineer a quasi-2D active system consisting of solid micron-sized light-absorbent platelets immersed in a thermotropic nematic LC. The platelets self-propel in the presence of light while self-inducing a localized NI phase transition. The sample's temperature, light intensity, and confinement determine three different regimes: a 2D large regime where the platelet-isotropic phase bubble is static and the NI interface remains stable; a compact motile-2D regime where the NI interface lies closer to the platelet's contour; and a motile-3D-confinement regime characterized by the emergence of multipolar configurations of the LC. We perform continuum-theory simulations that predict stationary platelet-LC states when confined in 3D. Our study in an intrinsically far-from-equilibrium landscape is crucial for designing simple synthetic systems that contribute to our understanding of harnessing liquid crystals' phase transitions to propel colloidal inclusions and trigger tunable topological reconfigurations leading to photonic responses.

## Introduction

Synthetic active materials, also known as micro/nanomotors, can harvest energy from their environment and transform it into mechanical forces for self-propulsion.[1,2] These materials have potential applications in transport and manipulation at microscopic length scales, such as drug delivery, local triggering of enzymatic systems, and water purification.[3,4] Extensive research has been done on synthetic active particles, including the self-propulsion of colloids driven by different phoretic mechanisms,[5] such as the ones triggered by catalytic reactions,[6] acoustic fields,[7–9] electromagnetic fields,[10] or light beams.[11–16] In addition, the development of self-propulsive materials has taken advantage of the structural properties of liquid crystals (LCs), such as the elastic interactions, the emergence of geometry-driven topological defects, and optical responses, giving rise to light-controlled,[17] chemically reactive,[18] or electrically and magnetically responsive systems.[19,20] Liquid crystals also enable the mobility of colloids through the induction of macroscopic thermal gradients to trap particles at the nematic-isotropic (NI) interface.[21,22] In recent years, experiments with LC emulsions have also shown different types of propulsive mechanisms.[23–27]

Nevertheless, the control of locally induced interactions at the NI interface has barely been exploited to design micromotors. In this context, we previously developed a system of light-activated spherical Janus particles within LCs.[28] In contrast, in this paper we want to explore the potentialities of geometry, shape, and confinement in an LC medium. We engineered a system of light-absorbing quasi-2D platelets suspended in the thermotropic nematic LC 4'-*Pentyl*-4-biphenylcarbonitrile (5CB). The platelets were formed by drying droplets of water-soluble and light-absorbing red dyes and propylene glycol on a glass surface. We studied the response of the platelets when confined in quasi-2D and 3D glass cells. In our design, the surrounding LC is indirectly heated when we illuminate the platelets with an LED source, controlling the light's output intensity. Unlike other approaches,[14,16,29] we used a structured liquid and a non-coherent light source. We first sought to change the temperature of the LC locally to trigger NI phase transitions with tunable interface sizes. With this approach, we found the conditions for the intensity of the applied light and sample temperature to trigger the platelet's self-propulsion confined in quasi-2D. We discovered that the close presence of the nematic-isotropic interface is crucial to activating the platelets, which would not happen otherwise. Finally, we tested how 3D confinement affected the evolution of the LC's configuration and the platelet's active motion. In a 3D space, the platelet moves in the horizontal plane and across the cell's height. In this case, we observe multipolar configurations of the LC while the platelet moves, tuned by the intensity of the applied light and, consequently, the local temperature. It has been previously shown that similar 5CB conical anchoring configurations appear on the surface of colloidal particles or glycerol droplets during the diffusion of surfactant molecules at the droplet-LC boundary,[30–32] in contrast to our case where the effect is maximized by the disk-like geometry of the platelet and the tunable light intensity. Lastly, we performed continuum theory simulations that predicted the distortion of the nematic director in steady state when the platelet is confined in 3D and self-propels while rising across the cell height. The simulations show the importance of the geometry restrictions in this rich platform of active inclusions within LCs, where shape and confinement lead to the emergence of tunable topological optical responses.

## Results and Discussion

As detailed in the Materials and Methods section, we placed tiny aqueous red food dye droplets onto OTS-functionalized glass substrates. After we dried the samples overnight, we observed the formation of solid quasi-2D platelets about 150–250 µm in diameter. Each individual platelet showed inhomogeneous patches due to the random local concentration of dry dye (see sketch in Figure 1a), with surface profilometry showing that the thickness of the platelet peaked mostly near the center (Figure 1b). We calculated a maximum platelet height of about 8 µm on average. We used Mylar films of 12 µm as glass spacers to confine the platelets in quasi-2D, and we used films of 100 µm for 3D confinement. After injecting 5CB between glass slides, we sealed the cell with epoxy resin (Figure 1a). $\hat{\boldsymbol{n}}$ is the nematic director, which accounts for the homeotropic LC anchoring. We used a dual top-bottom heating stage handled with a high-precision temperature controller to stabilize the sample's global temperature. We then observed the sample under an optical polarized microscope (see diagram, Figure 1c). Two LED sources simultaneously illuminated the sample. The microscope's white light source was used in transmission mode to image the experiments, while an external LED engine (peaked at λ ~540 nm) was used in reflection mode to activate the platelets. The activation light beam was collimated to ensure homogeneous illumination of the sample, and a custom cube consisting of a dichroic mirror and a filter increased the illumination efficiency and avoided camera saturation (see Materials and Methods for details). We measured the light absorption spectrum of the red food dye (Figure 1d) and its overlap with the LED illumination (a shaded-gray area in Figure 1d with a peak at λ ~540 nm), as well as the transmission of the dichroic mirror and filter. Recordings were obtained with two cameras, one connected to the microscope and the other placed at the eyepiece to cover the entire microscope's field of view (not shown in the diagram of Figure 1c for simplicity). Observations were made with and without setting linear polarizers up, as exemplified in Figure 1c. The platelet's birefringent contour indicates the LC's local reorientation at its surface.

According to the confinement restrictions, we identified two main regimes, quasi-2D and 3D. Depending on the sample's temperature and the intensity of the applied light for activation, the study of the platelets in quasi-2D confinement can be divided into two regimes: a large 2D regime where a platelet-light-induced isotropic phase bubble is static, and the NI interface remains stable, and a compact motile-2D regime where a light-induced NI interface lies closer to the platelet's contour. In turn, the 3D confinement approach shows motile platelets characterized by the appearance of light-induced multipolar configurations of the nematic LC. The characteristics of each regime are discussed in the following subsections.

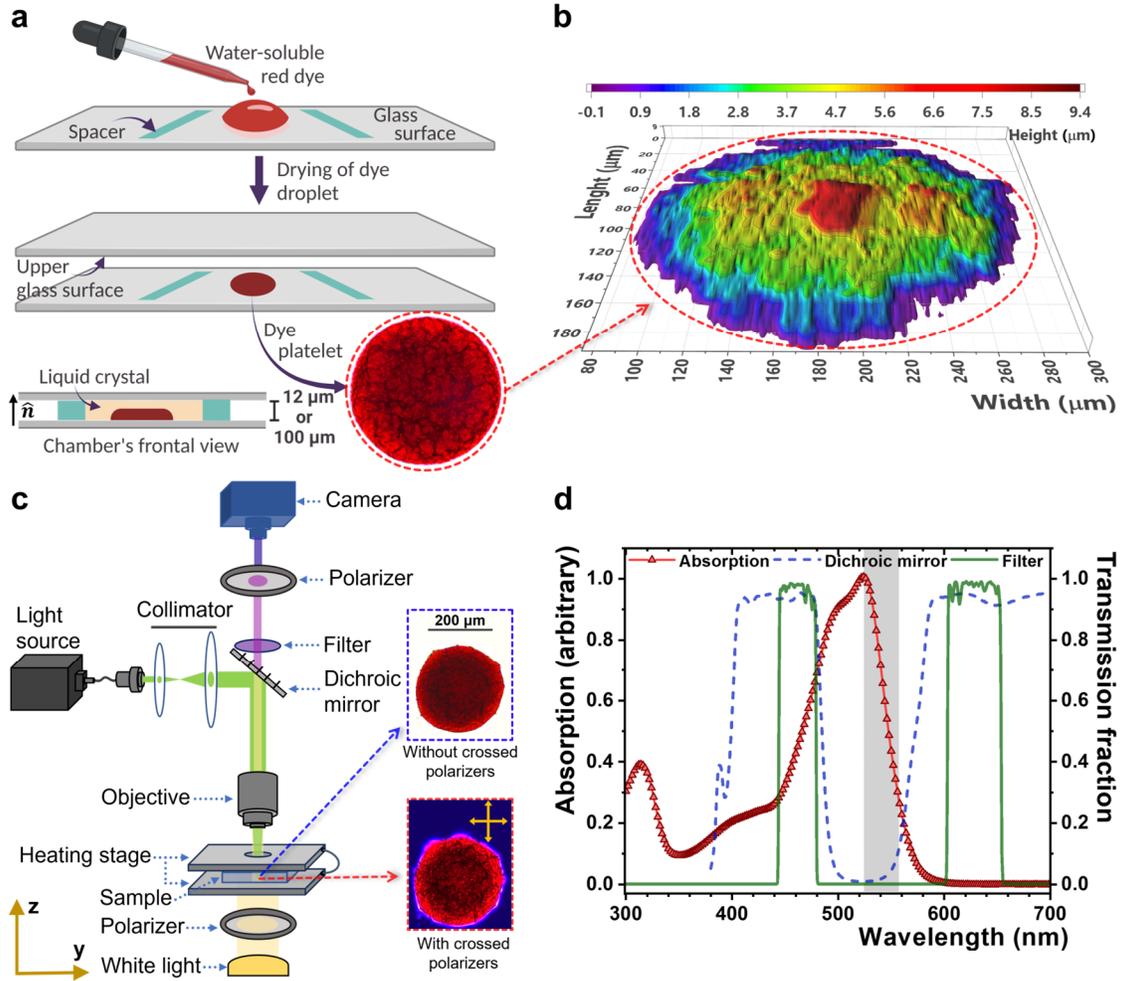

**Figure 1.** Schematics showing sample characteristics and experimental setup. a) Sample preparation process. The vector $\hat{n}$ is the nematic director, indicating the homeotropic LC anchoring. b) Profile of a platelet obtained using a profilometer. The color bar on the top indicates the height of each section. c) Experimental setup. All the elements are coupled to an optical polarized microscope. The inset micrographs correspond to a platelet immersed in the LC, observed with and without crossed polarizers. d) Absorption spectrum of the food dye and transmission fraction for the optical filters used in the experiments. The left *y*-axis corresponds to the dye absorption spectrum (red triangles). In contrast, the right *y*-axis corresponds to the transmission fraction of the dichroic mirror and filter (blue-dashed and green lines, respectively). The gray area indicates the spectral width of the LED that we use to activate the sample.

## *1. Platelet-isotropic bubble in quasi-2D confinement*

Our platelet-LC system reaches a global NI phase transition at a temperature $T_t$ = 35.2 °C. This transition temperature is captured at a 0.1 °C min$^{-1}$ rate using the heating stage. On the other hand, a local phase transition is induced at a lower global temperature, $T_g$, controlled with the heating stage while illuminating the sample with the green LED. As diagrammed in Figure 2a, an NI interface is formed with the isotropic region radially growing from the center of the platelet, with its size tuned by the light intensity. The size of the light-induced isotropic bubble depends on the size of the platelet and $T_g$. However, the heating response does not vary considerably for platelet diameters between 150 and 250 µm. For these experiments, we varied $T_g$ such that $\Delta T = T_t - T_g = 0.1$ °C, 0.25 °C, 0.5 °C, 0.75 °C, and 1 °C. At each global temperature, we varied the power of the light source over a range of 10% to 100% of the maximum, where the maximum corresponds to 125.06 ± 0.06 mW after considering light loss, with a tunability resolution of 1%. The LED system illuminates the entire field of view of the microscope's objective at the focal plane. The LED's wavelength used here does not directly affect the 5CB's temperature, which mostly absorbs UV light and scatters the rest, making it white to the naked eye while in the nematic phase and transparent in the isotropic phase. The green color light is absorbed primarily by the red dyes in the platelets. Although the red dye mixture absorbs light of a wide range of wavelengths, it mainly absorbs in the green region, consistent with our LED source (Figure 1d). Since the 10× objective's field of view area is 4.02±0.02 mm$^2$, the maximum irradiance achievable by the light source with a 100% output power on a platelet is $I = 31.11\pm0.17$ mW mm$^{-2}$. Owing to the LC homeotropic anchoring, changes in the sample were observed only when $T_t$ was surpassed during illumination and placed between crossed polarizers.

The first noticeable response is the vanishing of the platelet's birefringent contour, followed by the formation of a NI-interface ring surrounding the platelet. Then, a bubble of the isotropic phase rapidly grows radially from the center of the platelet to a stationary state (Figure 2b). As we expected, the closer $T_g$ is to $T_t$, the less light power is needed to increase the size of the NI interface. To quantify how the irradiated light and $T_g$ induce the NI phase transition on the sample, we plot the ratio ($A_t/A_p$) of the isotropic phase area to the platelet area as a function of the irradiance at different $T_g$ isotherms (Figure 2c). A non-linear response with the same trend is found in all cases; the micrographs in Figure 2c show the changes in a sample at constant temperature ($\Delta T = 0.25$ °C) but with different irradiance percentages (upper row) and at a constant irradiance (100%) but with different $T_g$ (right-hand column). Therefore, the same response ($A_t/A_p$) can be achieved by varying either $T_g$ or $I$.

We compared the effects of increasing $T_g$ using only the heating stage, and when illuminating the platelet with light (with global temperature control using the heating stage). In both samples, the cell thickness was 12 µm, and the LC was homeotropically anchored. The phase transition goes from nematic to isotropic and back to the nematic phase. The real-time response (Supporting Information: SI, Mov1 and Mov2) shows that the birefringent interface was induced when $T_t$ was globally or locally reached. In the first case, the NI phase transition nucleates along the sample with no control, and it takes about 10 s to get the whole sample in the isotropic phase when heating at 0.3 °C min$^{-1}$ rate. In the second case, $I$ was set to 100% and $\Delta T = 0.1$ °C; when illuminating the platelet, the response takes place within the temporal resolution of the camera (0.033 s). In addition, we observed that 3 s are sufficient for the entire field of view to transition to the isotropic phase. Once the light is turned off, the LC needs about 30–60 s to relax completely. As detailed in the following subsections, we observed the existence of two different platelet-isotropic bubble regimes in quasi-2D confinement determined by $A_t/A_p > 2$ (static regime, Figure 2b and 2c) and $0 < A_t/A_p \leq 2$ (motile regime, Figure 2c and 2d). Therefore, this ratio helps predict the platelet's ability to move – that is, the motility of platelets is triggered only for certain combinations of $T_g$ and light intensities.

*1.1. Large platelet-bubble regime*

This regime is subjected to values of $A_t/A_p > 2$ (see Figure 2b and 2c). By keeping a constant light power and global temperature $T_g$, we maintain the coexistence of both nematic and isotropic LC phases with no detectable fluctuations at the NI interface, as observed in SI, Mov3. In steady state, the enthalpy of the system equals the thermal energy, preventing the isotropic phase from becoming larger. Furthermore, the platelet displays negligible motility and remains at the center of the isotropic bubble (see Figure 3f to observe the platelet's sub-diffusivity within the large isotropic bubble), meaning that the system's symmetry is preserved with no net force on the platelet, which is discussed in the next subsection.

To quantify the thermal response of 5CB when it is locally heated up, we calculated temperature profiles (SI, Equation S3) across the isotropic phase, from the platelet contour to the NI interface, by solving the heat equation ($\nabla^2 T = 0$, Equation S1) in the stationary state for small Péclet numbers $Pe \ll 1$ ($Pe \propto V \alpha^{-1}$, $V$ is the speed of the platelet and $\alpha$ is the liquid crystal thermal diffusion coefficient) when conduction effects dominate the temperature field near the platelet. We approximated the temperature of the platelet using experimental snapshots at different light intensities (for detailed procedures, see SI and Figure S1). As a first simple approach, we considered a linear dependence between the energy absorbed on the platelet and the heat released in the LC. Figure S2a compares the temperature profile under different light intensities at $\Delta T = 0.25$ °C, while Figure S2b shows a diagram complementary to Figure 2c in which the characteristic parameter is $r_t/r_p$, with $r_t$ corresponding to the isotropic phase and $r_p$ to the platelet's radii as a function of the light power supplied to the platelet ($IA_p$). We also calculated the heat flux from the platelet contour to the NI interface using the differential form of Fourier's law (Equation S4) and obtained the heat flux as a function of light power $IA_p$ for all the studied $T_g$ (see Figure S2c).

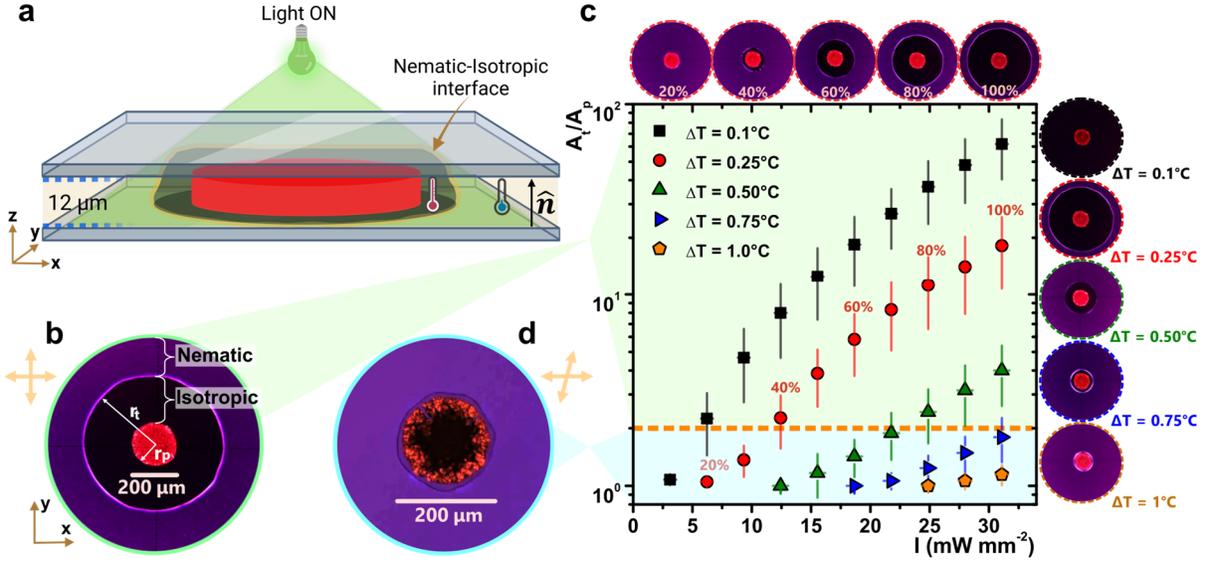

**Figure 2.** Local heating process and behavior of the self-induced nematic-isotropic transition. a) Schematic representation of the platelet-isotropic bubble in quasi-2D confinement. The global temperature controller and heating stage are not shown for simplicity. b) Experimental snapshot of the characteristic steady-state of the platelet and the NI interface when the system is illuminated with light for $A_t/A_p > 2$ (see diagram in c). The radii $r_t$ and $r_p$ are shown, corresponding to the isotropic phase and platelet respectively. The sample is placed in between crossed polarizers. c) Ratio of isotropic area induced around a platelet and platelet area in terms of the irradiance. The orange-dashed line marks the value $A_t/A_p = 2$. The error bars consider the standard error over experiments calculated for samples with slightly different platelet sizes. $\Delta T$ indicates temperature below $T_t \sim 35.2$ °C. The percentage of the total LED power is shown for the experiment with $\Delta T = 0.25$ °C. Micrographs at top are taken at constant temperature ($\Delta T = 0.25$ °C) and different irradiance percentages; micrographs at right are taken at constant irradiance (100%) and different $T_g$. d) Experimental snapshot of the characteristic state of a platelet and the self-triggered NI interface for the case of $0 < A_t/A_p \leq 2$. The sample is placed in between polarizers with one of them rotated to facilitate observation of the NI interface.

Figure S2c shows a non-linear decrease in flux in all cases, with more pronounced slopes for smaller $T_g$ or larger $\Delta T$. In contrast, the platelet contour temperature as a function of light power $IA_p$ shows an apparent linear monotonic increase (Figure S2d).

*1.2.Compact platelet-bubble regime*

From experimental observations, we learned that the platelet-isotropic bubble motility could only be triggered when the NI interface remains close enough to the platelet contour. Otherwise, a platelet immersed in an isotropic liquid with a symmetric medium at the surroundings would not experience a net force. This compact platelet-bubble regime corresponds to $0 < A_t/A_p \leq 2$, the lowest region in Figure 2c with corresponding snapshot in Figure 2d. Recordings of four representative cases show the displacement of compact platelet-isotropic bubbles over time (see selected initial and final frames, Figure 3a to 3d) and the corresponding speeds. The recordings combined with the platelet tracking data can be found in SI (Mov4 for Figure 3a, Mov5 for Figure 3b, Mov6 for Figure 3c, and Mov7 for Figure 3d). In all cases, $T_g$ was constant ($\Delta T = 0.5$ °C), and all 5CB molecules were in the nematic phase at $t = 0$. It takes a few milliseconds to melt the 5CB locally into the isotropic phase when the entire field of view is illuminated with green light. In three of the four cases shown, we held the light power constant at 25%, 33%, and 35% (Figure 3a to 3c). In the fourth case, we varied the power dynamically between 0% and 25% (Figure 3d). By tracking the platelet-isotropic bubbles' trajectories and speeds, we found that the platelets (cases a, b, and, to some extent, d) rotate within the isotropic bubble. The platelet's rotation affects the terminal speed, and higher terminal speeds can then be reached with lower power values when platelets do not significantly rotate. When we used a constant light power, the speed was approximately constant, regardless of rotation. However, the total kinetic energy is divided for platelets that rotate (cases a and b), decreasing the linear speed by an average of around 0.15 µm s$^{-1}$, compared to case c, which presented an average speed close to 0.5 µm s$^{-1}$. In the fourth case (d), we observed variations in the speed as the light power varied, although higher speeds were reached when the light power was lower and the NI interface was closer to the platelet boundary. All maximum speeds were approximately 0.4 µm s$^{-1}$ when the light power percentage was 18% and 20%, and approximately 0.2 µm s$^{-1}$ when the light power was 22% and 25%. Unlike the situation in the large platelet-bubble regime, in the compact platelet-bubble regime the NI interface shows fluctuations (see Supporting movies), likely triggered by the uneven distribution of dye crystals on the platelets' surface, which creates different hot spots that melt the LC. Figure 2d shows an experimental picture where the non-smooth NI interface is observed. Consequently, the motility process can be described as follows (see Figure 3e for schematic illustrations). First, at the steady state and before illumination, the LC's degenerate planar anchoring on the platelet boundaries keeps it in its original position (Figure 3e-1, and see Figure S3 for experimental reference when a λ-plate is used to observe LC-platelet anchoring). Second, a symmetry-breaking mechanism is switched on once the light is turned on, with the dye patches acting as hot spots (Figure 3e-2). Hence, the platelet-isotropic bubble experiences a net force, a momentum imbalance is generated, and the system moves until reaching a terminal speed. In the absence of light, the platelet stops moving after traveling a certain distance (Figure 3e-3). The impulse given to the platelets is a response to a local disequilibrium that emerges by the platelet itself when the temperature increases and induces a phase transition. This situation is unlike that in other experiments, where colloids remain in a nematic phase and move under focused or global thermophoretic effects[29,33] or are dragged by the NI interface when temperature gradients are globally controlled.[21]

A body of research has been carried out to explain the emergence of changes in direction and rotations in active systems. This approach considers gradients of viscosity that lead to *viscotaxis* in self-propelled systems, determining the dynamics and swimming direction. Furthermore, it has been suggested that living microorganisms perform viscotaxis to adapt to an inhospitable environment, with their physiology playing an important role in determining the swimming gait based on hydrodynamics. For instance, pullers such as bacteria like *Spiroplasma*[34] move faster up viscosity gradients (positive viscotaxis), while pushers such as *Escherichia coli*[35] swim better down the viscosity gradients (negative viscotaxis). Inspired by these biosystems, some theoretical models have been developed where effective propulsion forces are independent of viscous forces (frictional forces from Stokes' law) that determine the swimmer's directionality. Liebchen et al.[36] classify swimmers based on their body shape, assembling individual spherical beads to create rigid bodies affected differently by the orientation of their main axis and the effective propulsion force within the viscosity gradient field. In their model, only non-uniaxial linear swimmers (linearity in this context means that the propulsion force points towards the hydrodynamic center of mass of the swimmer) or any non-linear swimmers experience viscous torques to align with the viscosity gradients and then exert viscotaxis, due to a systematic mismatch of viscous forces acting on different body parts. Consequently, in this model under a viscosity gradient, an individual circular particle will not rotate and change its initial direction of motion unless the linearity is lost or the particle's body is formed by sites with different constituents that experience uneven drag forces, which is the closest scenario to our platelets. However, it does not consider a radial viscosity gradient from the particle itself. On the other hand, Datt and Elfring proposed a squirmer model with a prescribed surface velocity that sorts swimmers based on their swimming style (thrust) induced by the swimming gait.[37] Unlike the previous model, here the viscosity gradient couples the force to the bead's angular velocity and the torque to the respective translational velocity, which is not observed in a Newtonian fluid with uniform viscosity. They tested the case of a hot squirmer where viscosity radially increased from the bead,[38] similar to our system. However, in this scenario, the squirmer swims more slowly than a particle in a Newtonian fluid with uniform viscosity, and the angular velocity is zero because of the symmetry. Despite the advances in describing the phenomena, our system does not fully lie within the scope of these models. Although the platelets have a symmetric circular shape, their constituent crystallized dyes are randomly aggregated, which permits the uneven heating of the 5CB around them and induces the coexistence of two different LC phases. As explained above, the asymmetry is noticeable only when the NI interface remains close to the platelet, triggering translation with an as yet undefined thrust mechanism. Because the platelet is a rigid body and the forces are not pointing toward its geometric center (loss of linearity), it also experiences a net torque and consequently rotates. Nevertheless, the triggering mechanism and the emergence of a viscosity gradient are coupled phenomena in our case, which is not accounted for in either of the models.

To extend our understanding, we computed the diffusivity of the active platelets at $\Delta T = 0.5$ °C and different light powers. We also computed the values for platelets within the nematic phase at $\Delta T = 0.5$ °C, within the isotropic phase at a temperature 0.5 °C above the transition temperature, and the regime of a large isotropic bubble ($\Delta T = 0.5$ °C, 100% light power). The mean-squared displacement (MSD) is shown and compared in Figure 3f with the corresponding experimental snapshots. The active platelets behave almost ballistically at all light powers, which is a fingerprint of the activity. At the same time, the platelets in the other cases barely move, showing sub-diffusive behavior, even within viscosity gradients in the case of the large isotropic bubble regime. This proves that the symmetry-breaking mechanism created by the platelet conformation is not detected in this regime. The presence of a soft but sharp interface between a close-to-Newtonian fluid (LC isotropic phase) and an ordered viscoelastic fluid (LC nematic phase) close to the platelet alters its state. Therefore, the motility cannot be explained solely by considering only the smooth viscosity gradients. When the isotropic phase surrounds the platelet, an energetic barrier prevents the platelet from surpassing the NI interface, with elastic and interfacial tension contributions.[39] Some other approaches that may shed light on our system concern experiments on living swimmers[40] or theoretical models for gliders with different geometries close to and through sharp interfaces with different viscosities.[41] We hope our observations motivate researchers to study these out-of-equilibrium LC systems more thoroughly.

## *2. Platelet-bubble regime in 3D confinement*

After characterizing our system by keeping the platelet in quasi-2D confinement, we wondered what would happen if the platelet was confined in 3D. We postulated that the platelet could move and possibly rise owing to its disk-like geometry. These movements would manifest as an optical response induced by LC distortions. To test this idea, we prepared different samples following the described protocol using 100 µm thick spacers. We did not observe a dark isotropic region when illuminating the sample to induce the phase transition, likely due to the depth of the LC cell. However, the locally induced isotropic phase is identified as a transparent bubble with a birefringent ring around the platelet.

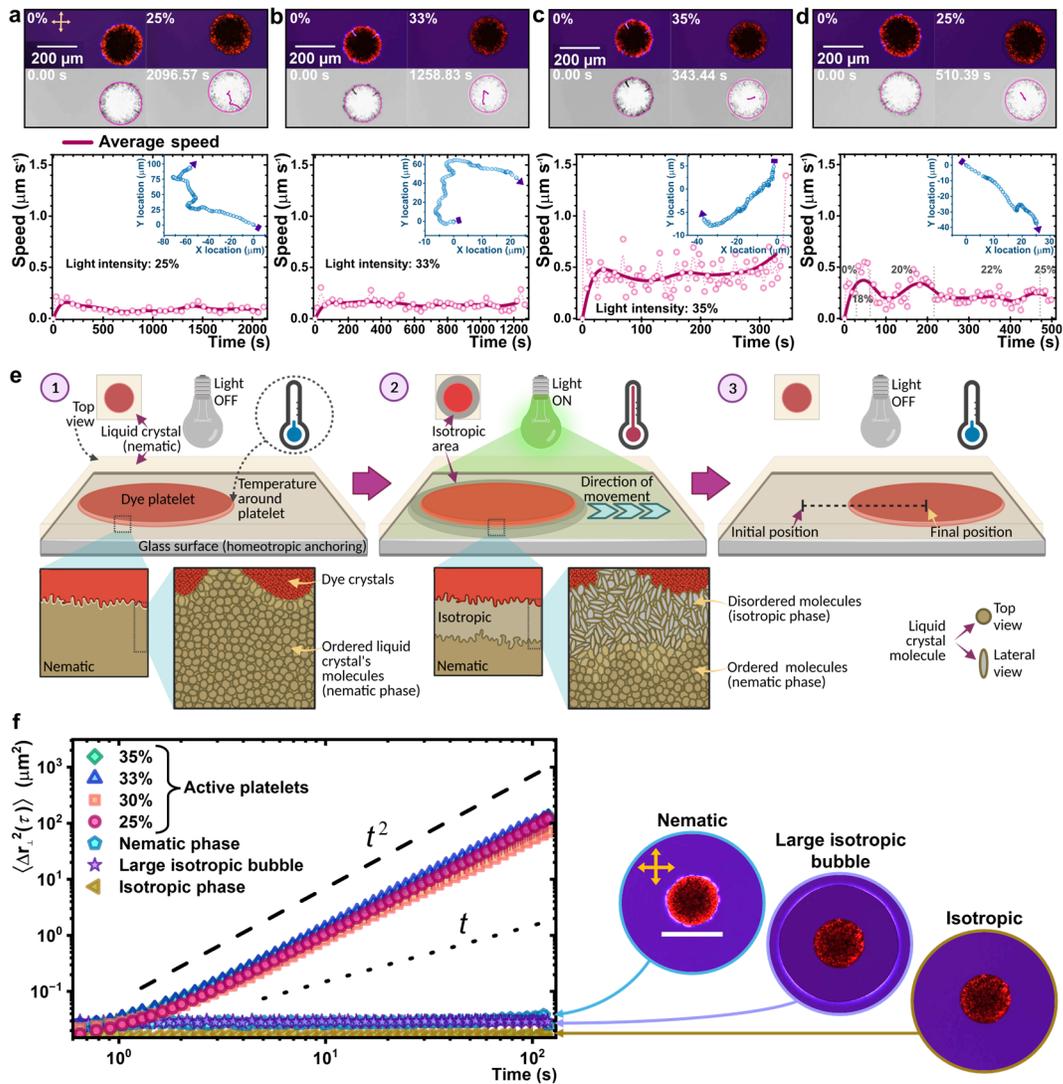

**Figure 3.** Displacement of platelets induced by light using a 12 µm cell. a-d) The top snapshots are cross-polarized optical microscopy images that show the platelets' initial and final positions (related movies can be found in the SI. $\Delta T = 0.5$ °C). The bottom pictures show the initial and final positions and the platelets' trajectory (magenta lines). Percentages of light power and total observation time are also indicated in each image. Corresponding plots below each snapshot show the platelets' speed as a time function. The average speeds are shown as a continuous line. a-c) are cases under constant light power; d) shows different light power as indicated in the plot. Insets are plots of total trajectories. The initial point is indicated with a purple rectangle, and the final position is shown with a purple arrow. e) Schematic of the process of light stimulation of a platelet embedded in 5CB and its motility response. The platelet contour yields a degenerated planar anchoring, while the glass treatment provides homeotropic anchoring. 1) With the light turned off, the platelet remains still, and the LC around it stays in the nematic phase. The LC molecules are ordered on the rough edge of the platelet since they are below $T_t$. 2) When the light is turned on, the local temperature around the platelet rises. $T_t$ is reached, and a non-birefringent area around the platelet is observed; the platelet-bubble glides forward if the isotropic-formed area is small. 3) When the light is turned off, 5CB rapidly recovers the globally controlled temperature, which prevents the platelet from moving. The process repeats if the light is turned on again. f) MSD plots for the active platelets as a function of light power and lag time at constant $\Delta T = 0.5$ °C. For direct comparison, MSD plots of the platelets within the nematic phase, isotropic phase, and the large-isotropic bubble regime are also shown, with corresponding snapshots on the right hand. The scale bar represents 200 µm. The dashed line shows the slope of the ballistic regime, while the dotted line represents the slope of the diffusive regime.

Figure 4 shows time-lapse optical microscopy images of a representative experiment that lasted for over an hour, in which the sample was kept at constant $\Delta T = 0.5$ °C. The investigation is divided into two parts; during the first part, the platelet encountered obstacles created by LC defects and disclination lines while moving (Figure 4a to 4f). During the second part, the platelet moves much more freely as it seems to be in an area with no obstacles (Figure 4h to 4m). The paths followed by the platelet during both parts of the experiment are shown in the inverted gray-scale pictures in Figure 4g and 4n, and the corresponding full sped-up movies can be found in the SI (Mov8, and Mov9). Figure 4o shows the entire trajectory and the speed changes. The sample shows a high optical response at a particular light power threshold. After the first 30 s of irradiation with a relatively low light power (35%), the platelet is quickly attracted to a defect ~90 µm away, with an instantaneous speed of over 1.5 µm s$^{-1}$ (average speed ~0.5 µm s$^{-1}$). During the attachment to the defect, the platelet generates a birefringent ring that is continuously evolving (Figure 4b and 4c). This ring is the nematic-isotropic interface, with the platelet remaining inside the isotropic phase bubble. Then, after the light power increases to 50%, two well-defined defects are created at the platelet-bubble's front and rear parts, allowing it to detach and continue moving. When located in a region free of obstacles, the platelet-bubble shows a monotonic speed increase from its lowest value when illuminated at 45% light power, to the maximum at 60%. At this point, the defect-pair redirects the platelet-bubble's pathway (Figure 4d) and prevents it from running aground towards other defects. The defect-pair rotates to overcome the obstacles and turn the platelet-bubble upwards. After ~3000 s, the platelet changes its apparent shape when passing over the remaining defects, appearing more ellipsoidal than circular. Then the platelet-bubble displays a different birefringent contour, and the speed drops to ~0.1 µm s$^{-1}$ (Figure 4e). At the end of the first part, the platelet escapes from the obstacle's area. In the second part of the experiment, the platelet finds a field that is free of obstacles. At constant light power of 65%, the system showed an LC-multipolar configuration in just a few seconds. Movies were analyzed to quantify the response times. The system's first response happened 0.07 s after the light was turned on. It took only 0.44 s to form birefringent rings displayed as a quadrupolar structure (Figure 4h). It took about 3.37 s for the quadrupole to distort and reconfigure into a 12-pole structure, fully formed after ~32 s (Figure 4i). While this evolution occurred, after about a minute, the platelet-bubble sped up and glided at a constant stable speed of ~0.4 µm s$^{-1}$, which was maintained for the rest of the experiment until the light was turned off. The multipole evolution is shown in Figure S4, which captures the formation in more detail. The platelet-bubble's path slightly bends during the motility process, probably due to long-range elastic interactions with LC defects and unperfect platelet shape.

However, these elastic penalties were not quantified. Even though the defect-pair is preserved and dictates the trajectory of the platelet-bubble, the platelet goes back rapidly to the first stages once the light is turned off, showing a transient quadrupolar optical configuration (Figure 4m). The platelet's shape becomes visually distorted into an ellipsoid when the octupole structure forms and goes back to the circular shape when the light is turned off (Figure 4, Figure S4 and Mov9 in SI). This change indicates the platelet being lifted from a side while moving towards the unlifted side, confirmed with microscope observations when different focal planes were found for opposite sites on the ellipsoidal-like platelet (see sketches of Figure 4p and Figure 4q). This effect can be explained by previous research where a levitation mechanism has been identified for colloids embedded in LCs.[42] This effect occurs when the LC anchors differently at the colloid and the substrate boundaries, bringing an incompatibility of the nematic director orientation in bulk with the orientation at the surfaces. In response, gravity is balanced by an elastic repulsion force near the walls, pushing the platelet up. In our scenario, the incompatibility is unstable when the NI interface is large enough to create a considerable elastic distortion of the nematic director at the platelet's neighborhood. Previous studies confirm the reorientation of the LC mesogens at the NI interface, corroborating our hypothesis.[43] The LC elastic response occurs unevenly, lifting only one side of the platelet. Moreover, the elastic distortion of the nematic director generates the flower-like birefringent multipolar pattern. We performed further experiments to explore this effect, complemented with computational continuum theory simulations (see Figure 5).

Figure 5a shows a series of experimental cross-polarized optical microscopy snapshots taken while we increase the illumination power while maintaining a constant temperature, $\Delta T = 0.7$ °C, below the NI transition temperature. The experiment consisted of switching on the illumination for 1 min and then 30 s without illumination, permitting the relaxation of LC. Each snapshot was taken close to completing the 1 min of illumination. The birefringence patterns evolve until showing a clear isotropic bubble around the platelet at 50% illumination power and then the emergence of multipolar flower-like patterns at 80-90% illumination power. Figure 5b shows essentially the same experiment as before but with a λ-plate placed between the crossed polarizers. The blue and red colors indicate the orientation of the LC mesogens (see diagram of Figure 5d). We can then infer the orientation of the nematic director at the birefringence flower-like patterns for different illumination powers. It is possible to observe the multipolar flower-like pattern from 80% to 100% illumination power, with intercalation of colors that indicate the twisting of the nematic director around the platelet.

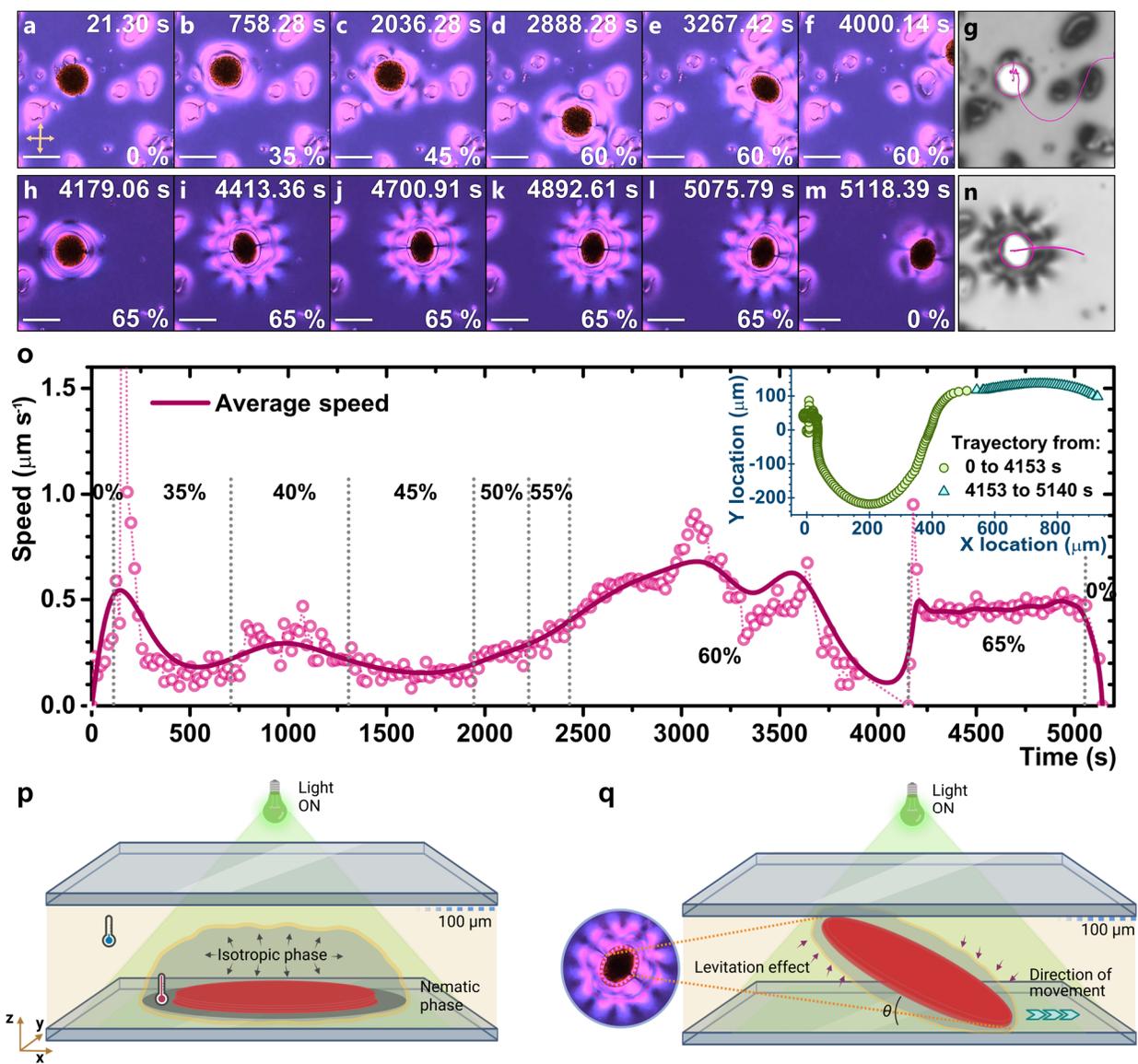

**Figure 4.** Displacement of a platelet induced by light and phase transitions in a 3D-confinement regime, where the cell is 100 µm thick and is kept at $\Delta T = 0.5$ °C. Cross-polarized optical-microscopy video frames a-f) show the first part of the investigation, while h-m) show the second part. g) and n) show the platelet's trajectory followed on each part of the experiment. The speed of the platelet is shown in the main plot o). The speed for the entire trajectory is represented with circles and a dotted line, while the speed for the average trajectory is shown as a continuous line. Percentages of total light power used at different time intervals are also indicated. The trajectory and distance the platelet traveled are shown in the inset plot, with circles indicating the first section of the experiment and triangles for the second part. The scale bar represents 200 µm. Corresponding movies can be found as SI. p) Schematic representations of the NI interface when the platelet is immersed in a sufficiently thick cell. q) The platelet lifts due to the incompatibility of the LC nematic director at the bulk, surfaces, and the NI interface. The elastic distortions of the LC at the NI interface permit the platelet to move smoothly and generate the flower-like birefringent pattern. The $\theta$ angle is shown as a reference for the platelet's inclination. Sketches are not to scale and are meant as references to clarify our discussion.

Corresponding movies (SI, Mov10, Mov11) show the dynamics of the platelet and LC during the experiment. When the platelet is activated, it moves from left to right, relative to the snapshots in Figure 5, and lifts at the rear side.

To obtain the equilibrium configurations of the LC in the presence of a platelet and explore the energetic landscape, we performed 3D numerical simulations employing an iterative finite-difference method and continuum mean-field Landau-de Gennes formalism.[44] The details of the simulations can be found in the Materials and Methods section. Briefly, the simulations were conducted on a platelet with degenerate planar anchoring surrounded by a disordered region simulating an isotropic phase within a nematic LC media with homeotropic anchoring at the top and bottom surfaces ($z$-direction). The free-energy landscape was examined by varying the parameter $\theta$ across different positions with the pivot point set to the right edge rather than the center of the platelet, where $\theta$ represents the angle between the platelet surface and the lower channel surface, as sketched in Figure 4q. We made variations every 5°, except for the maximum angle of 38°, permitted by the boundary conditions and size of the platelet, which is consistent with the experiments. Figure 5c shows top views of cross-polarized optical microscopy snapshots from simulations, and Figure 5d shows the corresponding ones simulating a λ-plate configuration. These pictures were obtained using the LCPOM Python package[45] from the Landau-de Gennes simulations. The birefringence lobes observed in the simulations do not capture the full optical response from the experiments. The differences can be attributed to the intrinsically far-from-equilibrium state of the experimental system. The simulations do not compute the constant energy input and the consequent temperature gradient from the platelet to the LC medium. Nonetheless, our simulations are a first approximation to understanding the phenomenom from the energetic point of view, with consistent results. The colors observed are optical responses integrated along the simulated system's whole depth ($z$-direction). More details can be found in the Materials and Methods section. Although there is no input energy source, the steady state results from minimizing the free energy for different angle inclinations match considerably well with the experimental observations. Figure 5e shows transversal views for $\theta = 0°$, 20° and 38°. These pictures permit the appreciation of the nematic director's distortion with the platelet's presence within an LC-disordered layer. At a higher inclination, the system configuration is perceived as more symmetrical, which can be interpreted as a more stable configuration to balance gravity with elastic and surface interactions in the $z$-direction, as occurs with the experimentally observed levitation effect.[42] For a further observation of the nematic director distortion at an inclination of $\theta = 38°$, Figure 5f shows three top-view simulation snapshots with λ plate and the corresponding nematic director from cuts of the top, middle, and bottom of the simulation box, as can be seen from the 3D Figure 5g. As mentioned before, the colors shown in Figure 5f snapshots result from integrating the optical response along the depth of the simulation box. Consequently, the colors match the nematic director better at different depths. In the Supporting Information (Figure S5), corresponding top-view images with λ plate and nematic director (Figure S5a) and 3D configuration (Figure S5b) for $\theta = 0°$ permit to observe the asymmetry of the system and great distortion of the LC, making the platelet more favorable to lift given the energetic penalties. To further evaluate the most favorable energetic landscape, we compared multiple simulations to identify the configuration of the lowest total energy density, allowing us to delineate the free energy minima for an equilibrium configuration (see Figure 5h and 5i). A minimum of the total free energy density is found from an inclination angle $\theta = 25°$ to $35°$, consistent with the configuration observed in experiments at 70% light power and above (see Figure 5a and 5b). Slight differences are observed for the total surface free energy density contributions, including the top and bottom boundaries, platelet, and NI interface surfaces (inset of Figure 5h), where the minimum is reached at $\theta = 25°$. At this angle, the platelet's tilted orientation helps to reduce interactions between the LC anchored at the confining walls and the LC at the platelet's NI interface, accommodating a less distorted nematic director.[46][47] However, elastic contributions increase (see Figure 5i), suggesting that the LC molecules undergo a distortion due to a non-smooth transition from the original anchoring conditions. Warping and straining of the nematic field contribute to the higher elastic energy. On the other hand, the off-center pivot point significantly impacts the system because of its asymmetric nature. As the platelet rotates, its surface interacts quite differently with the LC director, creating pronounced gradients in molecular alignment and thus increasing the elastic energy. In summary, at around $\theta = 25°$, the molecular alignment near the surface is energetically favorable, reducing the surface free energy. Nevertheless, the elastic free energy increases as the overall configuration needs bulk distortions.

In previous studies on LC multipolar configurations,[30–32] the optical responses were restricted to elastic quadrupole and hexapole structures driven by the conic anchoring of the mesogens on curved surfaces and tuned by the diffusion of surfactants within the LC medium. To directly compare with these observations, we also carried out continuum theory simulations featuring initial LC conic anchoring on the platelet surface without adding a surrounding disordered state layer. The results are shown in Figure S6 of the SI (Figure S6a and S6b for top view cross-polarized snapshots for different platelet inclination angles and colored images when simulated with λ plate. Figure S6c shows transversal views for $\theta = 0°$, $20°$ and $38°$. Figure S6d shows the simulation box's top, middle, and bottom layers with the nematic director at an inclination of 38°. Figure S6e is the corresponding 3D representation. Figure S6f and S6g are the total energy density, surface free energy density, and elastic free energy density plots). The birefringence lobes around the platelet are consistent, well-formed, and short-ranged, deriving in a well-behaved nematic director around the platelet. In contrast to the simulated case with an isotropic layer, the elastic and surface energies are about ten times lower here. Cross-polarized images then confirm that the presence of a non-evolving isotropic region does indeed facilitate the distortion of the nematic director and, consequently, the extension of the birefringence lobes into the surrounding area, fitting more closely with experimental observations. Finally, although we do not fully replicate the complete range of lobes observed experimentally, we observe tendencies that fit the experimental data.

## Conclusions

We engineered an active system of quasi-2D solid dyed platelets of ~200 µm, in which it was possible to trigger and locally tune a nematic-isotropic (NI) phase transition to propel the platelets, controlling the global temperature, $T_g$, while illuminating with an LED source having a wavelength of 542 nm. We analyzed the local phase transition and dynamics of the platelets within two scenarios: 2D and 3D confinements. In the 2D case, we observed two regimes with explicit behavior determined by the isotropic phase area to platelet area ratio, $A_t/A_p$. The large platelet-isotropic bubble regime is for $A_t/A_p > 2$, where the platelet barely moves, displaying sub-diffusive motion. We estimated temperature profiles and heat flux across the isotropic light-generated phase from experimental snapshots at different light power and constant-temperature steady states. In such states, the system's enthalpy equals the thermal energy, and a stable NI interface remains unchanged. If the light irradiation ceases, the bubble collapses. The second regime is the compact platelet-isotropic bubble for values $0 < A_t/A_p \leq 2$. Here, the platelet moves with a terminal speed that depends on the light power. Due to the drying process, the platelets experience an inhomogeneous distribution of dyes, rendering a relatively rough surface. Therefore, when the platelets are subjected to green light illumination, they heat up asymmetrically, inducing a microscopically uneven and fluctuating NI interface. This symmetry-breaking event ignites the platelet-bubbles' propulsion. We observed platelets' irregular trajectories, getting constant terminal speeds after a monotonic increase when the light power is constant. We also captured the platelets' diffusivity. The mean-squared displacement reveals that the active platelets have a ballistic behavior, unlike those not in the compact platelet-bubble regime that do not move significantly.

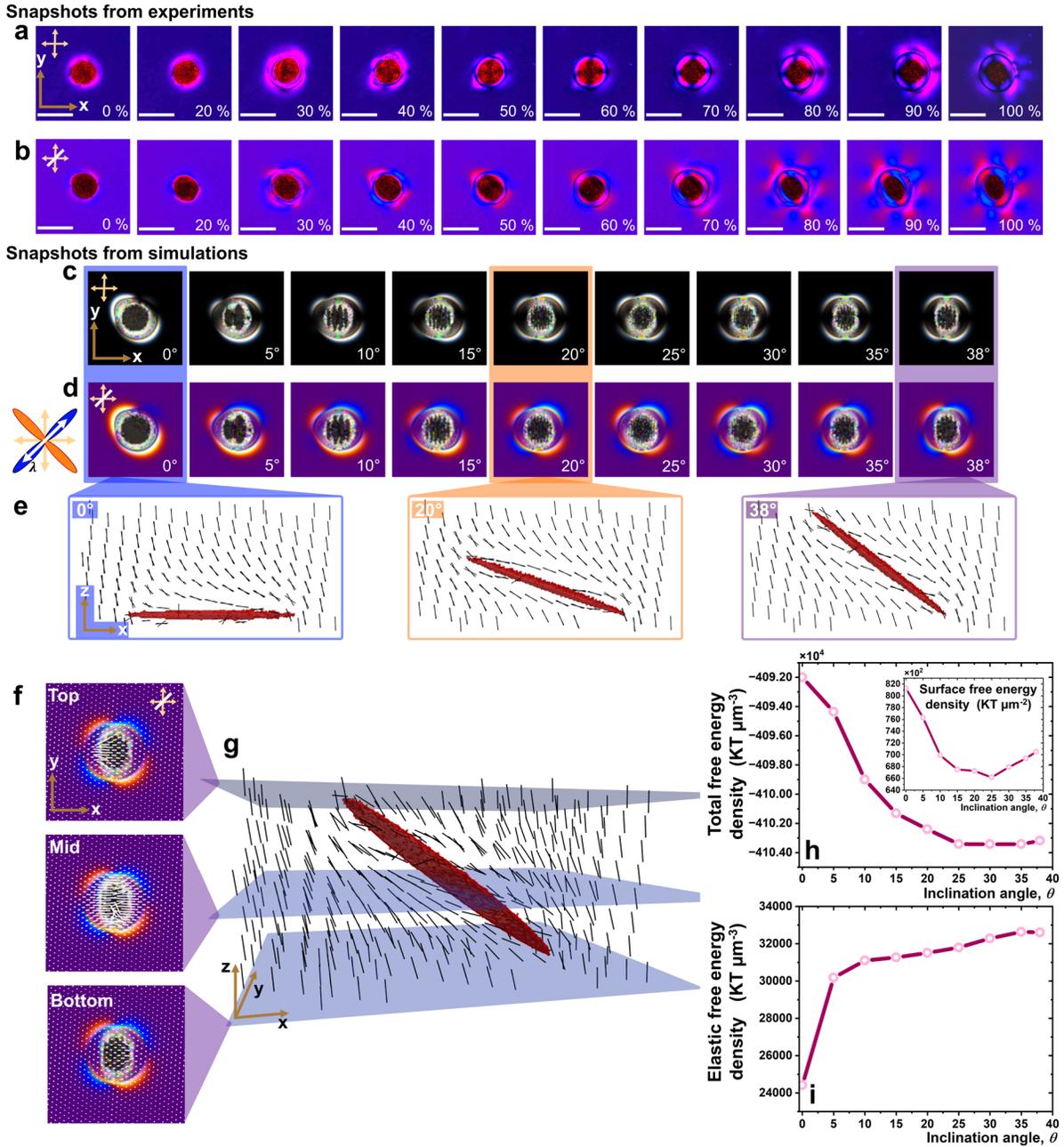

**Figure 5.** Comparison of experimental observations and continuum theory simulations with a disordered constant phase around the platelet for the 3D platelet-bubble confinement. a) Experimental cross-polarized optical microscopy snapshots. The light power varies while the temperature is fixed at $\Delta T = 0.7$ °C. b) Snapshots from experiments with the same conditions as before, with a λ-plate and crossed polarizers to show the direction of the mesogens for different light powers. c) Continuum theory simulation cross-polarized snapshots. Different inclination angles reproduce the experimental observations. d) Corresponding simulated snapshots with λ-plate and crossed polarizers. The colors show the direction of the mesogens, as pointed out in the diagram at the left. e) Cross-section of the simulation box displaying the $\theta = 0°$, $20°$ and $38°$. f) Simulated λ-plate of the top, middle, and bottom planes at the simulated inclination $\theta = 38°$. The nematic field is shown in white color lines. Since these are 2D images, the size of the director lines permits us to observe when they lie in the plane of the image or as dots when they point perpendicular from the plane. g) 3D snapshot of the platelet at $\theta = 38°$. The nematic field is shown as black lines. h) Total free energy density plot in terms of the inclination angle. The inset corresponds to the surface free energy density. i) Elastic free energy density plot in terms of the inclination angle.

Despite significant theoretical advances for explaining the viscotaxis of living systems within viscosity gradients using various models, there are still no complete models that describe our system in the presence of an out-of-equilibrium soft NI interface between a nonuniform viscous and ordered viscoelastic media. Lastly, we analyzed the 3D confinement scenario, where the local control of the LC temperature at a certain light irradiation threshold and the larger space above the platelet permitted the platelet to move while lifting unevenly, leading to different multipolar structures. As in the 2D confinement case, the LC around the platelet goes into the NI transition. When the platelet rises, a pair-defect is then formed at the NI interface at opposite regions of the platelet-isotropic bubble, which takes control of the swimming direction. The lift is explained by a levitation effect on colloids when the nematic director orientation in bulk is incompatible with its orientation at the surfaces. The incompatibility emerges at the NI interface, where the LC has a conical orientation instead of preserving the homeotropic direction of the cell boundaries. This unstable incompatibility creates a considerably uneven elastic distortion, lifting only one side of the platelet. We performed continuum theory simulations to quantify the energetic penalties from the LC distortions. At around a platelet inclination of 25°, the molecular alignment near the NI interface surface is energetically favorable, reducing the surface free energy. In contrast, the elastic free energy increases since the system's configuration needs bulk distortions compatible with the geometry and confinement restrictions. Simulated cross-polarized optical microscopy and λ-plate images using the novel LCPOM Python package permitted us to realize the importance of the geometry restrictions of the platelet shape that lead to the emergence of multipolar optical responses. Finally, the evolution of the observed LC multipolar structures resembles previously studied elastic dipole-quadrupole-octupole-hexadecapole transformations driven by LC interactions with colloidal particles or surfactants located on the surface of glycerol droplets within 5CB. In contrast, we obtained fast-responsive conical 12-pole configurations by tuning the light power and local temperature.

We have created a novel out-of-equilibrium active structure by combining an anisotropic 2D material within a structured fluid that can be subjected to locally tunable temperature changes through light. This is a rich platform for exploring different optical configurations due to LC driven by active colloids with different shapes and confinements. On the other hand, efforts to describe the motility mechanisms of living systems performing viscotaxis have been made in the past. However, current models do not accurately describe our system where the platelet activation and the trigger of a viscoelastic gradient are coupled. Furthermore, the coexistence of the nematic and isotropic phases permits the motility of the platelets. It would be worth exploring further to develop more complete models to describe systems like the one we propose.

## Materials and Methods

*Experimental details*

*Materials.* 4'-*Pentyl*-4-biphenylcarbonitrile (5CB) was purchased from Hebei Maison Chemical Co., Ltd (China). Red food dye was obtained from McCormick (Hunt Valley, MD) as a suspension that contains a mixture of Erythrosine, Allura Red AC, Propylene Glycol, and water with non-reported specific amounts. Sulfuric acid (95.0%-98.0%) was purchased from Sigma-Aldrich (US), hydrogen peroxide (30%) was purchased from Fisher Scientific (Pittsburgh, PA), and soda-lime glass microscope slides were purchased from Thermo Scientific (Portsmouth, NH). The glass slides were treated with Trichloro(*octyl*)silane (97%) (OTS) (Acros Organics, UK), heptane (Fisher Scientific, US), and ethanol (Sigma-Aldrich, US). Deionized water with a resistivity of 18.2 MΩ cm was obtained with a Milli-Q system (Millipore, Bedford, MA). Mylar film was purchased from Premier Lab Supply (Port St. Lucie, FL), and epoxy resin from Devcon (Danvers, MA). We used the AURA III engine (Lumencor, US) as the illumination source for activation, consisting of a set of 5 LEDs; in our experiments, we used the green channel (bandpass filter $\lambda = 542$ nm, FWHM = 33 nm. Maximum nominal output power = 500 mW). The light engine was connected to an upright Leica DM-2700P microscope (Germany) in transmission mode using a liquid light guide (see Figure 1c). Brightfield and linear polarized images and videos were obtained with a Leica MC170 HD camera. The camera recorded movies at 15 fps. To catch the whole objective's field of view, a cellphone with a 16 MP Sony built-in camera and a numerical aperture of f/2 was mounted to the microscope's eyepiece. This camera recorded movies at 30 fps. We used a N-Plan achromatic objective (10×). To increase the efficiency of the sample illumination and avoid saturation on both camera sensors, we used a custom filter cube (Chroma Technologies, US; dichroic mirror %T ≥ 95% for 470 nm ≤ $\lambda$ ≤ 480 nm, %R ≥ 95% for 495 nm ≤ $\lambda$ ≤ 545 nm, and transmission filter %T ≥ 95% for 445 nm ≤ $\lambda$ ≤ 470 nm and 605 nm ≤ $\lambda$ ≤ 650 nm). Light-power measurements were done using a Thorlabs PM160T power meter (US). Thermogravimetric measurements were done with a TGA instrument (TA Instruments, US). Platelets' profilometric scans were performed with a Bruker DektakXT-S profilometer (US). We used a dual top-bottom heating stage HCS402 to stabilize the LC temperature, controlled with a high precision temperature controller mk2000 (resolution of 0.001 °C; INSTEC, US).

*Sample preparation.* We built cells to confine the LC for experimental observations using glass slides. The slides were cleaned with piranha solution (70% $H_2SO_4$, 30% $H_2O_2$) and then rinsed profusely with deionized water and ethanol. The glass substrates were treated with OTS, as shown elsewhere[48] to provide a hydrophobic surface and induce LC homeotropic anchoring. On the other hand, the hydrophobic surface prevents the platelets from irreversibly adhering to the glass. Tiny droplets of red food dye were placed onto the functionalized glass substrates using a glass syringe (Hamilton, US), which determined the different platelet sizes, with diameters from ~150 µm to ~250 µm. The sample was then kept overnight at room temperature, allowing the water to evaporate. The presence of propylene glycol in the mixture increases the viscosity of the solution and induces the organization of inhomogeneous patches of dyes when the water evaporates (see Figure 1a). During the drying process, the droplets obtained their solid quasi-2D platelet shape, with a thickness shorter than 12 µm, the minimum cell thickness in our experiments (height to diameter ratio of ~1:25). Mylar films (12 and 100 µm) were used as glass spacers. 5CB was heated to 50 °C and then introduced gently into the glass cells by capillary effects. The cells were immediately sealed with 5-minute epoxy resin and allowed to cool down slowly to room temperature to favor an adequate nematic alignment. Profilometric scans were done to characterize the topography of the platelets. A series of 2D scans were run, each separated by 6 µm, covering a length of 400 µm with a resolution of 0.013 µm. 3D reconstructions of the 2D profiles were done with the Origin Pro software, where linear interpolation was done between individual 2D scans. Thermogravimetric measurements were obtained to estimate the content of solids in the dye solution and the approximate mass of an average-size platelet. Constant weight and isothermal (150 °C) measurements for different dye amounts were done, and a plot of the solids content *vs.* volume for each measure is shown in the Supporting Information (Figure S7). We estimated a mass of 0.07191±0.0032 µg for a 0.01 µL dye droplet. Platelets' tracking analysis was performed using the ImageJ software with the TrackMate plugin.[49]

*Experimental procedure.* Samples were allowed to reach thermal equilibrium before running the experiments. The system's global temperature ($T_g$) was increased from room to 32 °C at a rate of 0.5 °C min$^{-1}$. The rate was changed to 0.1 °C min$^{-1}$ until reaching a desirable temperature below the NI phase transition temperature $T_t$ = 35.2 °C. A white LED light source was used to illuminate the sample in transmission mode to image the experiments. Simultaneously, an LED source peaked at λ ~540 nm, illuminated the sample in reflection mode, for platelet activation. During the platelet activation, the dyes absorb the green color light and transform the excess optical energy into heat, increasing the temperature of the platelets. Which, in turn, self-triggers a NI phase transition that propels the platelet.

*Simulation details*

We employed the continuum mean-field Landau-de Gennes formalism[50] to obtain the equilibrium configurations of the confined liquid crystal in the presence of a platelet-like colloidal particle, as observed in our experiments. Within this framework, the free energy functional, $F$, is expressed in terms of the tensorial order parameter $\mathbf{Q}$. This parameter is defined as $\mathbf{Q}_{ij} = S\left(\mathbf{n}_i \mathbf{n}_j - 1/3\,\boldsymbol{\delta}_{ij}\right)$, with $i,j = 1, 2, 3$, and $\mathbf{n}_i$ being the $x$, $y$, and $z$ local director vector components and $S$ being the scalar order parameter.

The free energy functional accounts for the short-range ($f_{LDG}$), long-range elastic ($f_E$), and surface $f_S$ contributions and can be read as

$$F(\mathbf{Q}) = \int d^3 x \left[ f_{LDG}(\mathbf{Q}) + f_E(\mathbf{Q}) \right] + \int d^2 x f_S(\mathbf{Q}), \tag{1}$$

with,

$$f_{LDG} = \frac{A}{2}\left(1 - \frac{U}{3}\right)tr(\mathbf{Q}^2) - \frac{AU}{3}tr(\mathbf{Q}^3) + \frac{AU}{4}tr(\mathbf{Q}^2)^2, \tag{2}$$

where $U$ is the thermal parameter related to the isotropic-nematic transition and $A$ sets the energy density scale. The parameter $U$ determines the value of the scalar order parameter $S$ through

$$S = \frac{1}{4} + \frac{3}{4}\sqrt{1 - \frac{8}{3U}}. \tag{3}$$

The elastic contribution is given by

$$f_E = \frac{1}{2}\left[ L_1 \frac{\delta \mathbf{Q}_{ij}}{\delta x_k} \frac{\delta \mathbf{Q}_{ij}}{\delta x_k} + L_2 \frac{\delta \mathbf{Q}_{jk}}{\delta x_k} \frac{\delta \mathbf{Q}_{jl}}{\delta x_l} \right], \tag{4}$$

where $L_i$ are given by the Frank elastic constants for splay ($K_{11}$), twist ($K_{22}$), bend ($K_{33}$), and saddle-splay ($K_{24}$). It follows that $L_1 = \frac{1}{2S^2}\left[K_{22} + \frac{1}{3}(K_{33} - K_{11})\right]$ and $L_2 = \frac{1}{S^2}(K_{11} - K_{24})$.

The last term in Equation 1 considers the surface contribution to the free energy due to the channel and the platelet. In our experiments, we observed a ring around the platelet above a certain power of illumination, which can be attributed to the local increase in the particle's temperature, causing an isotropic corona. Another possibility is that such an increase in the platelet's temperature causes fluctuations in the preferred planar alignment at the particle's surface,[51] which can be modeled by a conical anchoring condition. Therefore, in our simulations, we impose a homeotropic anchoring at the channel's surface, while for the platelet, we consider both isotropic (I) and conical (C) anchoring conditions. Accordingly, in the first set of simulations, the surface free energy reads

$$f_S = f_{channel}^H + f_{platelet}^I. \tag{5}$$

For the second set of simulations, we have

$$f_S = f_{channel}^H + f_{platelet}^C. \tag{6}$$

The surface free energy associated with the homeotropic anchoring can be described using the Rapini-Papoular free energy,[52]

$$f_S^H = \frac{1}{2} W_H \left( \mathbf{Q} - \mathbf{Q}^0 \right)^2, \tag{7}$$

where $W_H$ is the homeotropic anchoring energy, and $\mathbf{Q}^0$ is the preferred surface tensor order parameter. For the isotropic anchoring, we consider a randomly oriented director field at the platelet's interface that does not evolve. While for the conic-degenerate anchoring, we use the expression proposed by Zhou et al.,[31]

$$f_S^C = W_C \left( \mathbf{P}'_{ik} \tilde{\mathbf{Q}}_{kl} \mathbf{P}'_{lj} - S_{eq} \cos^2 \theta_e \mathbf{P}'_{ij} \right)^2. \tag{8}$$

$\tilde{\mathbf{Q}}_{kl} = \mathbf{Q}_{kl} + 1/3\, S_{eq} \boldsymbol{\delta}_{kl}$, where $S_{eq}$ is the order parameter (Equation 3) in equilibrium. Here, $W_C$ is the strength for the conic-degenerate anchoring, $\mathbf{P}'_{ij}$ is a projection operator given by $\mathbf{P}'_{ij} = v_i v_j$ with $v$ the normal vector to the surface, and $\theta_e$ the polar angle normal to the surface.

The total free energy is calculated by using a finite-difference scheme over a cubic grid with a resolution of 7.15 nm, which is equal to the coherence length of 5CB. Minimization is achieved by the iterative Ginzburg-Landau relaxation method, where $\mathbf{Q}$ evolves to equilibrium via

$$\frac{\delta \mathbf{Q}}{\delta t} = -\frac{1}{\gamma} \left[ \mathbf{\Pi} \left( \frac{\delta F}{\delta \mathbf{Q}} \right) \right], \tag{9}$$

with boundary conditions such that $\mathbf{\Pi}\left[(\delta F/\delta \nabla \mathbf{Q})\cdot v\right] = 0$. The parameter $\gamma$ represents the diffusion coefficient and the operator $\mathbf{\Pi}(\mathbf{B}) = 1/2(\mathbf{B} + \mathbf{B}^T) - 1/3\,tr(\mathbf{B})\mathbf{I}$ guarantees a symmetric and traceless **Q**-tensor.

We used the following numerical parameters. The platelet radii are $R_x = R_y = 429$ nm and $R_z = 21.45$ nm; the channel dimensions, with periodic boundary conditions along the $x$ and $y$ axes, are $L_x = L_y = 2860$ nm and $L_z = 600.6$ nm. The functional parameters are $A \approx 1 \times 10^5$ J m$^3$ and $K_{11} \approx 5 - 7 \times 10^{-12}$ N.[53] The other elastic moduli, $K_{22}$, $K_{33}$, and $K_{24}$ can be set as the splay modulus. We use $U = 3.5$, corresponding to a bulk scalar order parameter $S = 0.62$. We set the elastic constants as $L_1 = 0.769$ and $L_2 = 1.537$. $W_{channel} \approx 4.2 \times 10^{-4}$ J m$^{-2}$ and $W_{platelet} \approx 6.3 \times 10^{-4}$ J m$^{-2}$ are the channel homeotropic anchoring and platelet surface strengths, respectively, associated with $f_s$ in Equation 5.

We generated simulated cross-polarized optical microscopy (POM) and lambda-plate POM images using the LCPOM Python package[45] from the minimized configurations taken from Landau-de Gennes simulations. LCPOM is based on the Ondris-Crawford method for producing the brightness profile as light travels across a nematic liquid. For calculations, the system is discretized into layers, and the propagation of monochromatic light is modeled by the multiplication of the Jones matrix of each layer that accounts for the retardation caused by the nematic order. In order to generate the colored images, a combination of several calculations with different wavelengths is done and combined to produce the RGB values for human vision. Additionally, the used wavelength distribution is taken to match the source light from experiments. In this work, we used 20 wavelengths and modeled the refractive indices from a three-band model.[54]

## Supporting Information

Supporting Information is available at the end of the document.
The movie description is included in the Supporting Information.
Movies are available upon request to tavera@uchicago.edu and atv.tavera@gmail.com

## Author Contributions

A. T.-V. and J. J. d. P. conceived the work. A. T.-V. and D. M.-S. performed experiments and analyzed the data. V. N. M. shared the experimental facilities for pilot experiments and discussed the experimental data analysis. A. T.-V. and G. P.-L. performed theoretical calculations. G. P.-L., O. E. R.-L., J. A. M.-G., and J. J. d. P. conceived and performed numerical simulations. A. T.-V., D. M.-

S., G. P.-L., and J. J. d. P. wrote the paper. All the authors revised the manuscript, made improvements, and agreed on its final version.

## Acknowledgments

A. T.-V. gratefully acknowledges Prof. Jan Lagerwall and Prof. Teresa Lopez-Leon for useful discussions and suggestions. The authors thank Chuqiao Chen for supporting the use of the LCPOM Python package to obtain colored simulation images. Likewise, all the authors thank Prof. Stuart Rowan for sharing their polarized optical microscope, which permitted us to perform the experiments. This work made use of the Searle Cleanroom at the University of Chicago, funded through Award Number C06RR028629 from the National Institutes of Health–National Center For Research Resources. Instrumentation was procured with funding generously provided by The Searle Funds at The Chicago Community Trust (Grant A2010-03222). This work was also supported by the University of Chicago Materials Research Science and Engineering Center, which is funded by the National Science Foundation under award number DMR-2011854, by the Army Research Office through the MURI program under award no. W911NF13-1-0383 and by the Harvard University Materials Research Science and Engineering Center, which is funded by the National Science Foundation under award number DMR-2011754.

## Conflict of interests

The authors declare no conflict of interests.

# Supporting Information

## Light-Activated Motion, Geometry- and Confinement-Induced Optical Effects of 2D Platelets in a Nematic Liquid Crystal

Antonio Tavera-Vázquez, Danai Montalvan-Sorrosa, Gustavo Perez-Lemus,
Otilio E. Rodriguez-Lopez, Jose A. Martinez-Gonzalez,
Vinothan N. Manoharan, Juan J. de Pablo

**Associated movie descriptions can be found at the end of the document.**

**Temperature profile and heat flux estimation from experiments**

The heat equation in the stationary state is solved ($\nabla^2 T = 0$), which corresponds to the experiment with constant light power and controlled $T_g$ with a nematic-isotropic (NI) interface in steady state. A simple approximation was considered with an ideal cylindrical-shaped platelet. Any heat transfer through the boundaries was neglected. The heat equation, solved in cylindrical coordinates, is

$$\frac{1}{r}\frac{\partial}{\partial r}\left(\kappa r \frac{\partial T}{\partial r}\right) + \frac{1}{r^2}\frac{\partial}{\partial \theta}\left(\kappa \frac{\partial T}{\partial \theta}\right) + \frac{\partial}{\partial z}\left(\kappa \frac{\partial T}{\partial z}\right) = 0. \tag{S1}$$

The boundary conditions are

$$\begin{aligned} T(r) &= T_p \quad &\text{at} \quad & r = r_p, \\ T(r) &= T_t \quad & \quad & r = r_t, \end{aligned} \tag{S2}$$

with $T_p$ the temperature at the contour of the platelet of radius $r_p$ and $T_t$ the temperature at the NI interface with isotropic-phase radius $r_t$ (see Figure 2b). Given the azimuthal symmetry and the null heat transfer through the boundaries in the $z$-direction, the Equation S1 is reduced to a one-dimensional problem in the radial direction. The solution of Equation S1 is

$$T(r) = \frac{\delta T \ln(r/r_t)}{\ln(r_p/r_t)} + T_t, \tag{S3}$$

with $\delta T = T_p - T_t$. $T(r)$ values are subjected to the input values of $r$ restricted by the boundary conditions. $T_t = 35.20 \pm 0.01$ °C, $r_p$, and $r_t$ are known from experiments.

$T_p$ was determined using a series of experimental observations at each constant global temperature and tunable light power as a thermometer. Figure S1 illustrates the case for $\Delta T = 0.25$ °C. At 0% light power, the platelet remains at $T_g$. When shining with 20% light power, the energy absorbed on the platelet is released in the LC as heat in the form $Q_{20\%} \propto T_{20\%} - T_g$. Precise experimental observations showed that this amount of light was sufficient to melt the LC and induce the NI interface right at the contour of the platelet. Within thermal equilibrium, $T_{20\%} = T_p = T_t$, therefore $Q_{20\%} \propto \Delta T$. Increasing the light power by an additional 20% leads to a heat release of $Q_{40\%} \propto T_{40\%} - T_{20\%}$. As a first approximation, we considered a linear dependence between the energy absorbed on the platelet and the heat released in the LC. Then, it is possible to conclude that $Q_{20\%} = Q_{40\%}$. If the process is repeated, adding the same amount of light each time, then $Q_{20\%} = Q_{40\%} = Q_{60\%} = Q_{80\%} = Q_{100\%}$. After solving the last equations, $T_p$ is calculated in terms of known quantities; $T_{40\%} = \Delta T + T_{20\%}$, $T_{60\%} = 2\Delta T + T_{20\%}$, $T_{80\%} = 3\Delta T + T_{20\%}$ and $T_{100\%} = 4\Delta T + T_{20\%}$. The same approach was used for all the other studied cases. In all of them, $T_p$ is a function of $\Delta T$.

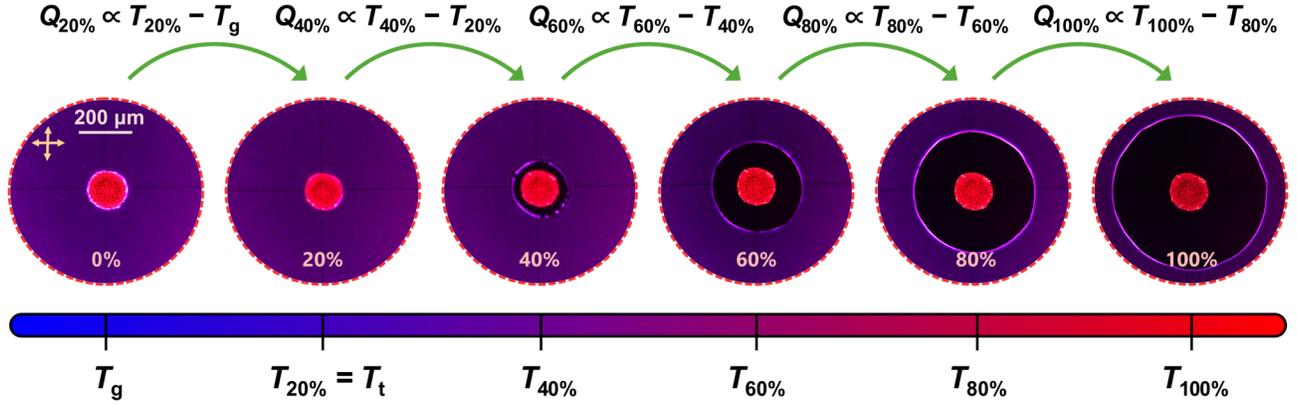

**Figure S1.** Estimation of the platelet temperature, $T_p$. A thermometer was created with a scale dictated by the configuration of the NI interface when shined at different light powers and constant $T_g$.

The heat flux along the isotropic phase was calculated using the differential form of Fourier's law $J = \kappa \nabla T$. In our simplified model, the temperature gradient is the derivative of the temperature profile (Equation S3) in the $r$ direction. The analytic form of the heat flux is then

$$J(r) = -\frac{\kappa \delta T}{r \ln(r_p/r_t)}, \tag{S4}$$

in which we have added a negative sign for consistency with the experiment and the boundary conditions in Equation S2. $\kappa = 1.507 \pm 0.001$ mW cm$^{-1}$ °C$^{-1}$ is the thermal conductivity, and the value used in our calculations was obtained from previous experiments.[S1]

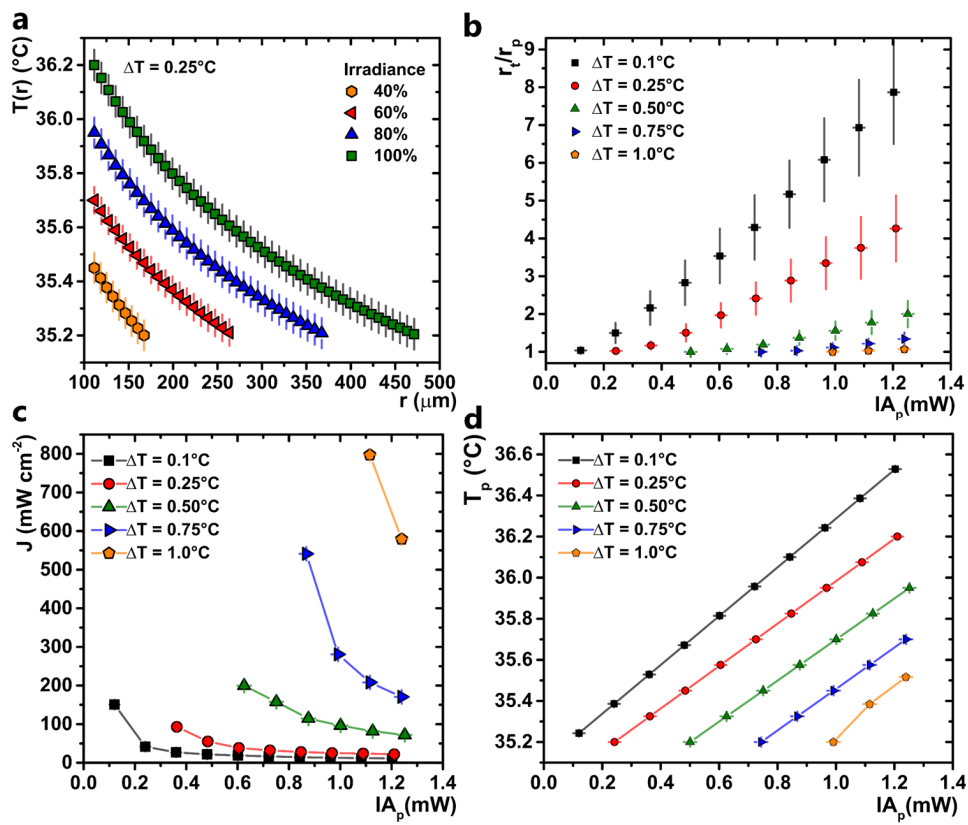

**Figure S2.** Results after solving the heat equation using experimental values. a) Calculated temperature profiles in the isotropic phase for the case of $\Delta T = 0.25$ °C with boundaries at the platelet's contour and the NI interface. b) Isotropic phase radius to platelet's radius ratio as a function of the light power supplied to the platelet, $IA_p$, when considering a circular shape. c) Heat flux as a function of $IA_p$. d) Change of the platelet's-contour temperature, $T_p$, as a function of $IA_p$. Different colors and figures represent the results for different global temperatures.

**Liquid Crystal anchoring at the LC-platelet interface**

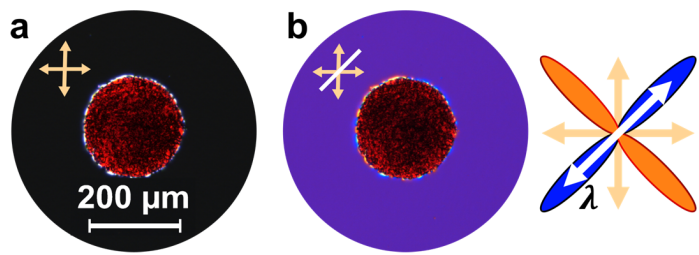

**Figure S3.** Cross-polarized microscopy images of a platelet in 5CB at $\Delta T = 0.25$ °C. a) The platelet shows birefringence at its border due to the anchoring of LC molecules. b) A λ-plate was set up to show the direction of the anchored mesogens at the platelet contour. The direction of the λ-plate with respect to the crossed polarizers is schematized with a diagonal white line on top of the crossed arrows (snapshot) or as a double-sided white arrow on the sketch that also shows colored mesogens, with the colors in correspondence to the orientation of their main axis according to the snapshots.

**Evolution of conic LC-multipole configurations in 3D confinement**

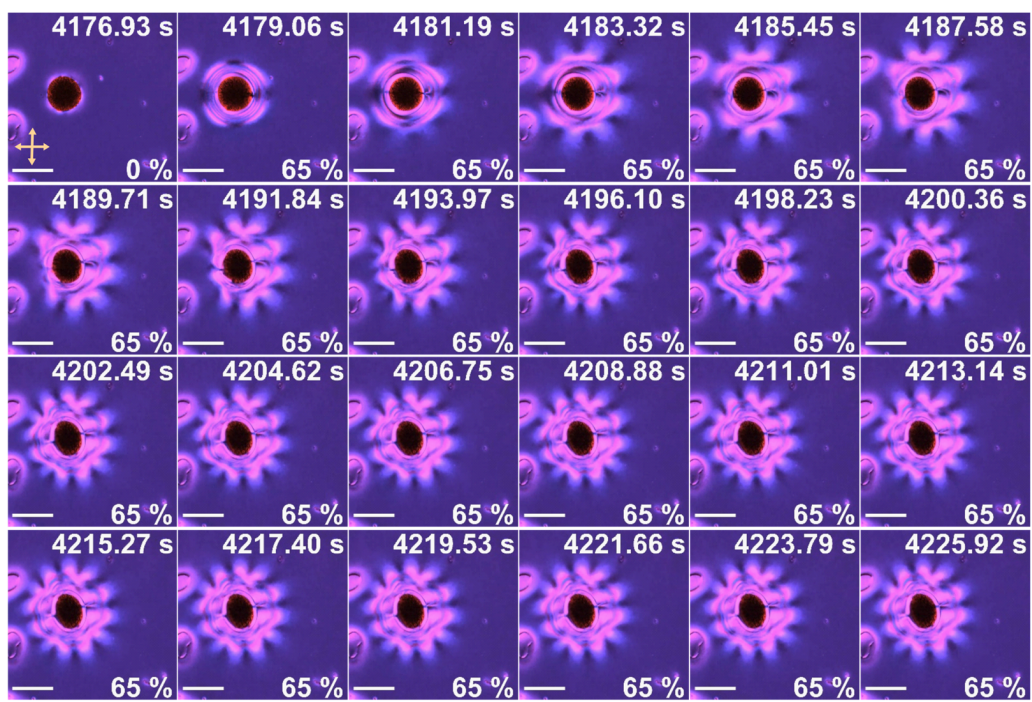

**Figure S4.** Snapshots of the platelet's evolution when forming the flower-like shape birefringent pattern in the case of a cell 100 µm thick. It takes ~32 s from starting the illumination to establish the final state of the platelet. Each snapshot was taken every 2.13 s. The scale bar represents 200 µm.

**Continuum theory simulations**

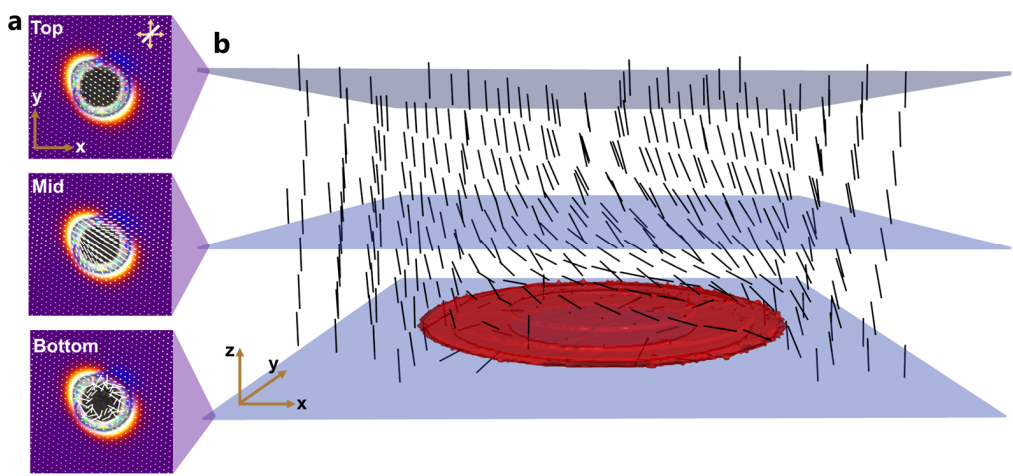

**Figure S5.** Continuum theory simulations with a disordered constant phase around the platelet for the 3D platelet-bubble confinement. a) Simulated λ-plate of the top, middle, and bottom planes at the simulated inclination $\theta = 0°$. The nematic field is shown in white color. Since these are 2D images, the size of the director lines permits us to observe when they lie in the plane of the image or as dots when they point perpendicular from the plane. b) 3D snapshot of the platelet at $\theta = 0°$. The nematic field is shown as black lines.

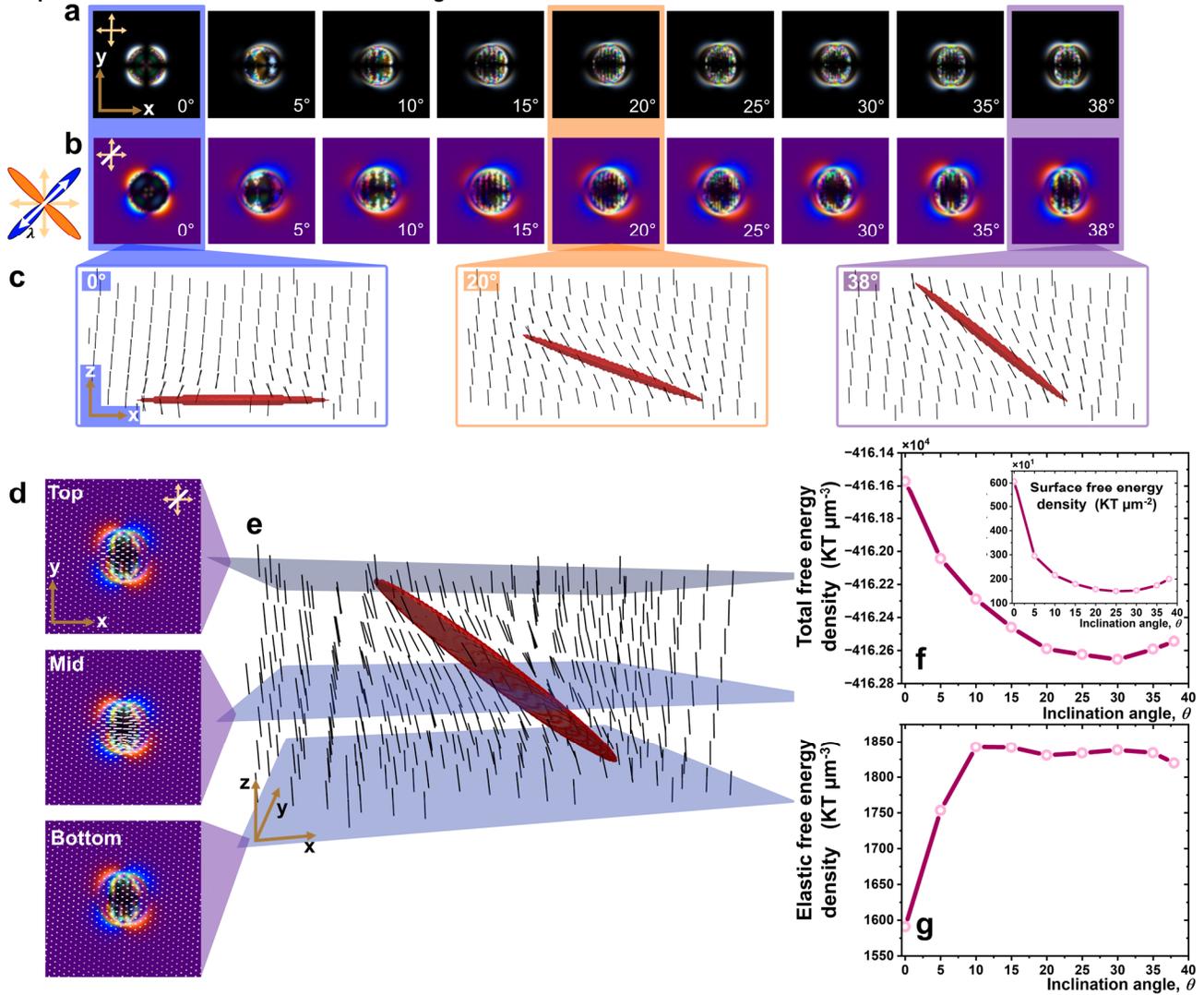

**Figure S6.** Continuum theory simulations with a conic anchoring around the platelet for the 3D platelet-bubble confinement. a) Cross-polarized snapshots. Different inclination angles reproduce the experimental observations. b) Corresponding simulated snapshots with λ-plate and crossed polarizers. The colors show the direction of the mesogens, as pointed out in the diagram at the left. c) Cross-section of the simulation box displaying the $\theta = 0°$, 20° and 38°. d) Simulated λ-plate of the top, middle, and bottom planes at the simulated inclination $\theta = 38°$. The nematic director is shown in white color. Since these are 2D images, the size of the director lines permits us to observe when they lie in the plane of the image or as dots when they point perpendicular from the plane. e) 3D snapshot of the platelet at $\theta = 38°$. The nematic director is shown as black lines. f) Total free energy density plot in terms of the inclination angle. The inset corresponds to the surface free energy density. g) Elastic free energy density plot in terms of the inclination angle.

**Platelet mass fraction calculation**

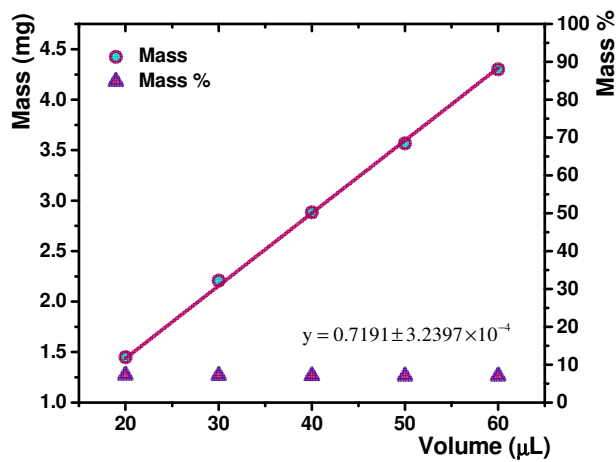

**Figure S7.** Platelet mass calculation *vs.* sample volume by TGA measurements. Mass values were obtained (circles) once the corresponding mass percentage values were constant and reached the minimum value at constant temperature = 150 °C (triangles). Linear regression permitted the estimation of a mass of 0.07191±0.0032 µg for a 0.01 µL dye droplet.

**Movie descriptions**

**Mov1.** Global nematic-isotropic phase transition of the 5CB nematic liquid crystal in the presence of a platelet. The whole sample is heated at a 0.3 °C min$^{-1}$ rate. The isotropic phase nucleates along the sample at the transition temperature $T_t$ = 35.2 °C. Once the field of view fully transitions into the isotropic phase, the sample remains there for about a minute to return to the nematic phase later. The sample is cooled at the same rate. The sample is placed between crossed polarizers, and the dichroic mirror and filter are set to compare directly with the active situations.

**Mov2.** Local nematic-isotropic phase transition of 5CB. The sample is kept at a constant temperature of $\Delta T$ = 0.1 °C below the transition temperature while illuminated at 100% light intensity. Due to the light absorbance by the platelet, the full field of view transitions into the isotropic phase within 3 s after illumination starts. A clear nematic-isotropic interface is observable which size depends on the illumination intensity. After 6.5 s, the light is turned off, and the sample rapidly returns to the nematic phase, as observed by the birefringence optical response. The sample is placed between crossed polarizers, and the dichroic mirror and filter are set.

**Mov3.** Local nematic-isotropic phase transition of 5CB in the large platelet-isotropic bubble regime. The sample is kept at a constant temperature of $\Delta T$ = 0.5 °C below the transition temperature while illuminated at 100% light intensity. The platelet displays negligible apparent motility and remains at the center of the isotropic bubble. The nematic-isotropic interface remains in a steady state while the light intensity is constant. The sample is placed between crossed polarizers, and the dichroic mirror and filter are set.

**Mov4.** Sample in the compact platelet-isotropic bubble regime, corresponding to the snapshots in Figure 3a. The sample is kept at a constant temperature of $\Delta T$ = 0.5 °C below the transition temperature while illuminated at 25% light intensity. The movie is sped up to capture the platelet's motion and trajectory. The left frame shows the experimental observation of the sample placed between crossed polarizers with dichroic mirror and filter. The right frame displays the sample in an inverted grayscale, tracking the platelet's trajectory.

**Mov5.** Sample in the compact platelet-isotropic bubble regime, corresponding to the snapshots in Figure 3b. The sample is kept at a constant temperature of $\Delta T = 0.5$ °C below the transition temperature while illuminated at 33% light intensity. The movie is sped up to capture the platelet's motion and trajectory. The left frame shows the experimental observation of the sample placed between crossed polarizers with dichroic mirror and filter. The right frame displays the sample in an inverted grayscale, tracking the platelet's trajectory.

**Mov6.** Sample in the compact platelet-isotropic bubble regime, corresponding to the snapshots in Figure 3c. The sample is kept at a constant temperature of $\Delta T = 0.5$ °C below the transition temperature while illuminated at 35% light intensity. The movie is sped up to capture the platelet's motion and trajectory. The left frame shows the experimental observation of the sample placed between crossed polarizers with dichroic mirror and filter. The right frame displays the sample in an inverted grayscale, tracking the platelet's trajectory.

**Mov7.** Sample in the compact platelet-isotropic bubble regime, corresponding to the snapshots in Figure 3d. The sample is kept at a constant temperature of $\Delta T = 0.5$ °C below the transition temperature while illuminated at varying increasing light intensities, such as 18%, 20%, 22% and 25%. The variation is displayed at the bottom right side of the left frame. The movie is sped up to capture the platelet's motion and trajectory. The left frame shows the experimental observation of the sample placed between crossed polarizers with dichroic mirror and filter. The right frame displays the sample in an inverted grayscale, tracking the platelet's trajectory.

**Mov8.** Sample in the platelet-isotropic bubble regime in 3D confinement, corresponding to Figure 4a to 4g snapshots. The platelet is confined in a 100 µm thick cell and is kept at $\Delta T = 0.5$ °C while illuminated at different increasing light intensities, from 35% to 60%, with changes every 5%, as shown in the bottom right side of the left frame. The field of view is full of topological defects. The platelet remains motionless until the activation light is turned on. The liquid crystal around the platelet rapidly melts and evolves while triggering the platelet's motion. During the first moments after illumination starts, the platelet is attracted to a topological defect and remains dynamically attached until 50% light intensity is reached. Here, topological defects are arranged around the platelet, located at the front and rear sides, leading the motion and direction of the platelet. A clear bubble surrounds the platelet, which we identify as the isotropic phase. At this stage, the platelet moves between and over the topological defects. After sufficient time under 60% intensity illumination, the platelet adopts an ellipsoidal shape and shows multipolar birefringence structures around it. The movie is sped up to capture the platelet's motion and trajectory. The left frame shows the experimental observation of the sample placed between crossed polarizers with dichroic mirror and filter. The right frame displays the sample in an inverted grayscale, tracking the platelet's trajectory.

**Mov9.** Sample in the platelet-isotropic bubble regime in 3D confinement, corresponding to Figure 4h to 4n and Figure S4 snapshots. The platelet is confined in a 100 µm thick cell and is kept at $\Delta T = 0.5$ °C while illuminated at a constant 65% light intensity. When the light is turned on, the platelet rapidly transitions state and develops an ellipsoidal shape with multipolar textures around it. These textures evolve from a quadrupolar to a 12-pole structure, fully formed after around 32 s. The platelet's ellipsoidal shape shows when the platelet lifts. During the first second of the movie, the focal plane is set to the bottom of the sample with a clear visualization of topological defects and front of the platelet. After that, the focal plane is shifted to where the rear of the platelet is now focused, and the rest of the platelet and topological defects are defocused. Once the light is turned off, the platelet returns to its original configuration. The movie is sped up to capture the platelet's motion and trajectory. The timer is set up to continue the time count from Mov8. The left frame shows the experimental observation of the sample placed between crossed polarizers with dichroic mirror and filter. The right frame displays the sample in an inverted grayscale, tracking the platelet's trajectory.

**Mov10.** Sample in the platelet-isotropic bubble regime in 3D confinement, corresponding to Figure 5a snapshots. The platelet is confined in a 100 µm thick cell. It is kept at $\Delta T = 0.7$ °C while illuminated at different 1 min intervals with increasing light intensities, from 10% to 100%, with changes every 10%, and intercalated intermittent 30 s intervals of light switched off as indicated in the movie. The optical textures evolve due to the light intensity and equilibrium time to reach a steady state. The movie is sped up to capture the platelet's motion and trajectory. The sample is placed between crossed polarizers with dichroic mirror and filter.

**Mov11.** Sample in the platelet-isotropic bubble regime in 3D confinement, corresponding to Figure 5b snapshots. The platelet is confined in a 100 µm thick cell. It is kept at $\Delta T = 0.7$ °C while illuminated at different 1 min intervals with increasing light intensities, from 10% to 100%, with changes every 10%, and intercalated intermittent 30 s intervals of light switched off as indicated in the movie. The optical textures evolve due to the light intensity and equilibrium time to reach a steady state. The movie is sped up to capture the platelet's motion and trajectory. The sample is placed between crossed polarizers with dichroic mirror, filter, and a λ-plate to capture the imaging light deflection that permits inferring the liquid crystal mesogens' direction.

**Supporting Information References**